\documentclass[%
 reprint,
 amsmath,amssymb,
 aps,
]{revtex4-2}

\usepackage[dvipdfmx]{graphicx}
\usepackage{dcolumn}
\usepackage{bm}
\usepackage[margin=20truemm]{geometry}
\usepackage[utf8]{inputenc}
\usepackage{color}
\usepackage{ulem}

\usepackage{comment}

\newcommand{\bra}{\left\langle}
\newcommand{\ket}{\right\rangle}

\newcommand{\der}[2]{\frac{d #1}{d  #2}}

\newcommand{\var}[2]{\frac{\delta #1}{\delta  #2}}

\newcommand{\bv}[1]{{\boldsymbol #1}}
\newcommand{\ep}{\epsilon}
\newcommand{\eq}{{\rm eq}}
\newcommand{\ex}{{\rm ex}}
\renewcommand{\ss}{{\rm ss}}
\newcommand{\eff}{{\rm eff}}

\newcommand{\tr}{\rm tr}
\newcommand{\est}{\rm est}

\newcommand{\nc}{n_{\rm c}}
\newcommand{\tep}{\tilde \epsilon}

\DeclareRobustCommand{\erase}{\bgroup\markoverwith{\textcolor{red}{\rule[.5ex]{2pt}{0.4pt}}}\ULon}
\newcommand{\red}{\textcolor{black}}

\begin{document}

\preprint{APS/123-QED}

\title{Divergent stiffness  of one-dimensional growing interfaces}

\author{Mutsumi Minoguchi}
\author{Shin-ichi Sasa}%
\affiliation{%
 Department of Physics, Kyoto University, Kyoto 606-8502, Japan
}%

\date{\today}

\begin{abstract}
  When a spatially localized stress is applied to a growing one-dimensional
  interface, the interface deforms. This deformation is described by
  the effective surface tension representing
  the stiffness of the interface. We present that the stiffness exhibits divergent behavior in the large system size limit for a growing interface with thermal noise, which has never been observed for equilibrium interfaces. Furthermore, by connecting the effective surface tension with a space-time correlation function, we elucidate the mechanism that anomalous dynamical fluctuations lead to divergent stiffness.
\end{abstract}

\maketitle

{\it Introduction.---}
The statistical behavior of many-interacting elements out of equilibrium has attracted attention for a wide range of systems \cite{Eisert2015, Julicher2018, KPZ_review_T2018}. A remarkable feature of such systems is that the standard relations in equilibrium systems no longer hold. For example, phase order in two dimensions is not observed for equilibrium systems at finite temperatures \cite{Mermin-Wagner, Hohenberg}, while it emerges for active matters \cite{Toner} or sheared systems \cite{Nakano}. The particular nature of out-of-equilibrium systems is not limited to phase transition problems. The phenomenon we study in this Letter is the singular response against a perturbation.


{In studying response properties of equilibrium systems,
the fluctuation response relation is useful. That is, 
the static response against a perturbation is connected to
static fluctuation properties in the system without perturbation. 
As a result of this relation, 
the response is found to be finite except for phase transition points because static fluctuations are normal
in general. In contrast, the static response against a perturbation imposed to a non-equilibrium steady state is not determined by static fluctuation properties. Although several expressions of the static
response for out-of-equilibrium systems have been proposed \cite{HaradaSasa2005, SpeckSeifert2006,Chetrite2008,Baiesi2009,Prost2009,Baiesi2013} and {experimentally studied \cite{Blickle2007, Solano2009,Toyabe2010}} for the last two decades,
the most primitive method is to consider the time evolution of
the perturbation \cite{Mclennan, Zubarev, Kubo1991}. This means that 
the dynamic properties of fluctuations influence the static response
if there is no special property such as a detailed balance condition.
Therefore, a singular response behavior can be observed without tuning
system conditions.}


{To demonstrate the singular response of many interacting elements out of equilibrium,}
we specifically study a one-dimensional interface, whose height is defined in $0 \le x \le L$. The interface deforms when a localized stress is applied. For equilibrium interfaces \cite{EW}, \red{which do not grow but fluctuate in an equilibrium environment,} their \red{mean profile} in the linear response regime is expressed by a quadratic function of $x$, where its curvature is determined by the surface tension $\kappa$. 
Now, let us consider growing interfaces \cite{KPZ}. \red{We can
 numerically confirm that }
the deformation against the weak localized stress is still described by a quadratic function of $x$. \red{In this case},  the curvature of the interface is characterized by the effective surface tension $\kappa_{\rm eff}$. \red{We then find
that $\kappa_{\rm eff}$ diverges  as $L \to \infty$.} In other words, growing interfaces exhibit divergent stiffness.


{ We attempt to {explore} the mechanism of the divergent stiffness by formulating a fluctuation-response relation.}
This problem is reminiscent of the standard linear response theory around an equilibrium state. For example, when considering heat conduction for a Hamiltonian system in contact with two heat baths with temperatures $T_1$ and $T_2$, $T_2-T_1$ is treated as a perturbation \cite{Dhar2008}. In this case, the linear response formula is the Green-Kubo formula, which expresses the conductivity in terms of the time integration of the current correlation function at equilibrium \cite{Kubo1991}. Similarly to heat conduction, we expect that the effective surface tension $\kappa_{\rm eff}$ can be expressed as the time integration of a certain time correlation function. In this Letter, we derive such a formula using a generalized fluctuation theorem associated with the excess entropy production.


Based on the response formula, we study the divergent stiffness. As is known, some low-dimensional systems exhibit an anomaly in the large-distance and long-time properties of the time correlation function \cite{Dhar2008}. In such systems, the decay rate of a time correlation function is so small that its time integration is not bounded in the large system size limit \cite{FNS, Dhar2008, Lepri2016}. By combining this property with the response formula,
{the mechanism of the divergent stiffness is understood.
We emphasize that the method we propose in this Letter can be applied
to other spatially extended systems out-of-equilibrium.}

{\it Setup.---}
The one-dimensional interface defined in
$0 \le x \le L$ is investigated. The height of the interface at time $t$ 
is expressed by $h(x,t)$, which is collectively denoted by $\hat h = (h(x))_{x=0}^L$.
For simplicity, the periodic boundary condition $h(0,t)=h(L,t)$ is assumed.
An external stress $ \ep p_{\ex}(x)$ is imposed on the interface, where
the total force $ \ep \int p_{\ex}(x) dx$ is set to zero to avoid the additional drift of the interface. 
{We first study an equilibrium interface.} 
The free energy of the interface
is assumed to be
\begin{equation}
  F^{\ep}(\hat h)\equiv \int_0^L dx
  \left[\frac{\kappa}{2}
    (\partial_x h)^2-\ep p_{\ex}(x)h(x) \right],
\end{equation}
where $\kappa$ represents the surface tension. 
The fluctuation properties are described by the 
following stochastic model \cite{EW}:
\red{
\begin{equation}
\partial_t h = -\frac{1}{\gamma}\var{F^\ep (\hat h)}{h} 
+\sqrt{\frac{2T}{\gamma}}
\xi,
\label{eq-int}
\end{equation}
}
where $\gamma$ is the dissipation constant; $T$ is the temperature of the
bath with the Boltzmann constant set to unity; $\xi$ is the Gaussian white 
noise satisfying
\begin{equation}
\bra \xi(x,t)\xi(x',t') \ket=\delta(x-x')\delta(t-t').
\end{equation}  
Thus, it is immediately confirmed that the expectation of the interface shape under the external stress is given by
\begin{equation}
\kappa \partial_x^2 \bra h(x) \ket_{\eq}^{\ep}+  \ep p_{\ex}(x)=0,
\label{eq-res}
\end{equation}
where $\bra \cdot \ket_{\eq}^{\ep}$ denotes the expectation in the equilibrium state of the system with the external stress $\ep p_\ex(x)$.
For simplicity, focus is placed on the case where 
$p_\ex(x)=\delta(x)-1/L$.
By solving (\ref{eq-res}) \cite{SM}, we obtain 
\begin{equation}
  \bra h(x)-h(0) \ket_{\eq}^{\ep} =\frac{\epsilon}{2 L \kappa}
  \left[ \left(x-\frac{L}{2} \right)^2-\frac{L^2}{4} \right].
\label{eq-shape}
\end{equation}

    \begin{figure}[t]
    \includegraphics[width=8cm]{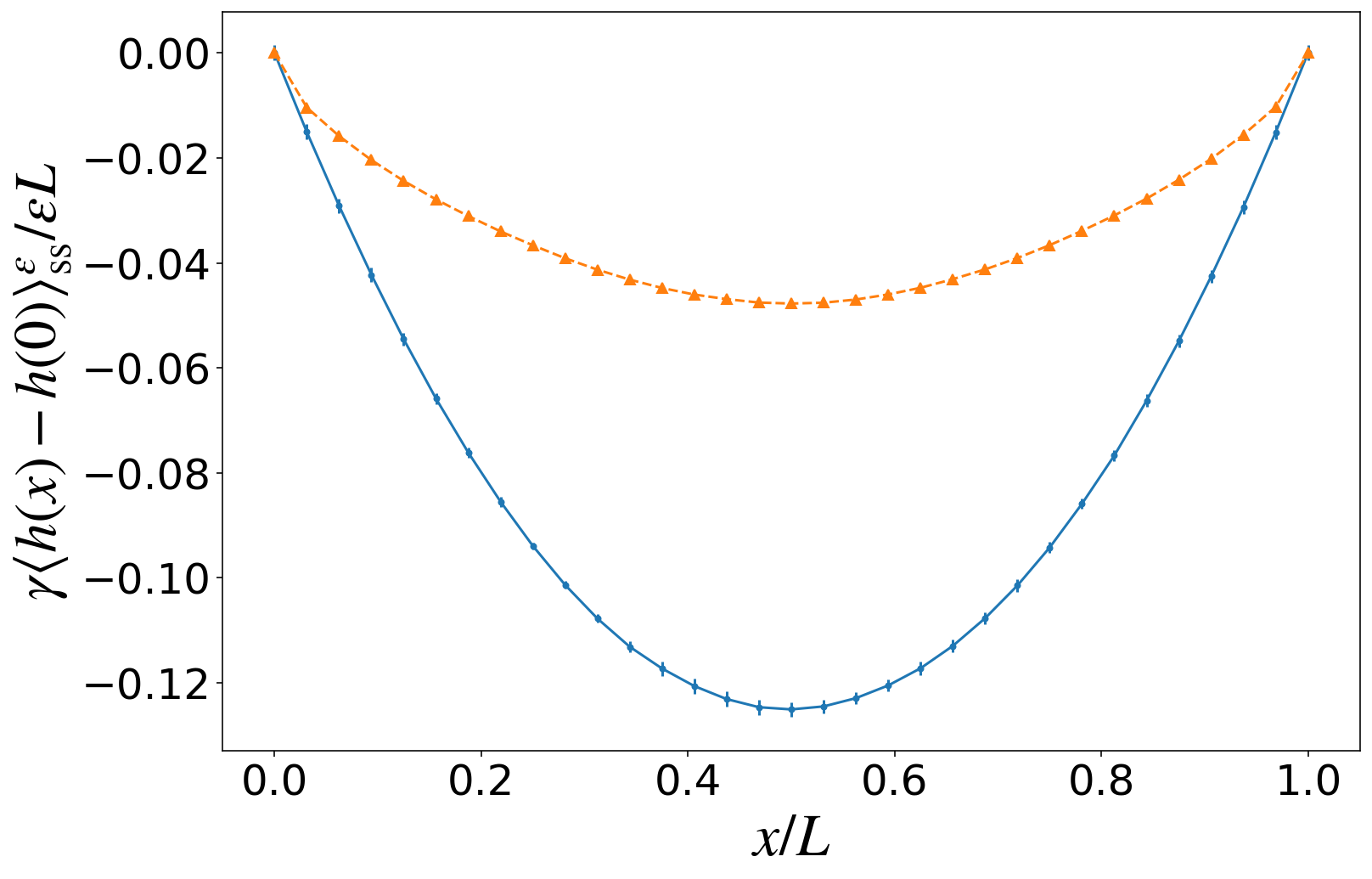}
    \caption{\label{fig:quad1} Time-averaged
      patterns in steady state under the external
      stress with $\epsilon/\gamma=0.01$. The system size is $L=16$.
      The curvature of the growing interface ($v_0=5$, {triangular-orange symbol})
      is smaller than that of the equilibrium interface ($v_0=0$, {round-blue symbol}). {The {symbols} are joined by lines for visual aid.}
 }
    \end{figure}

    \begin{figure}[t]
    \includegraphics[width=8cm]{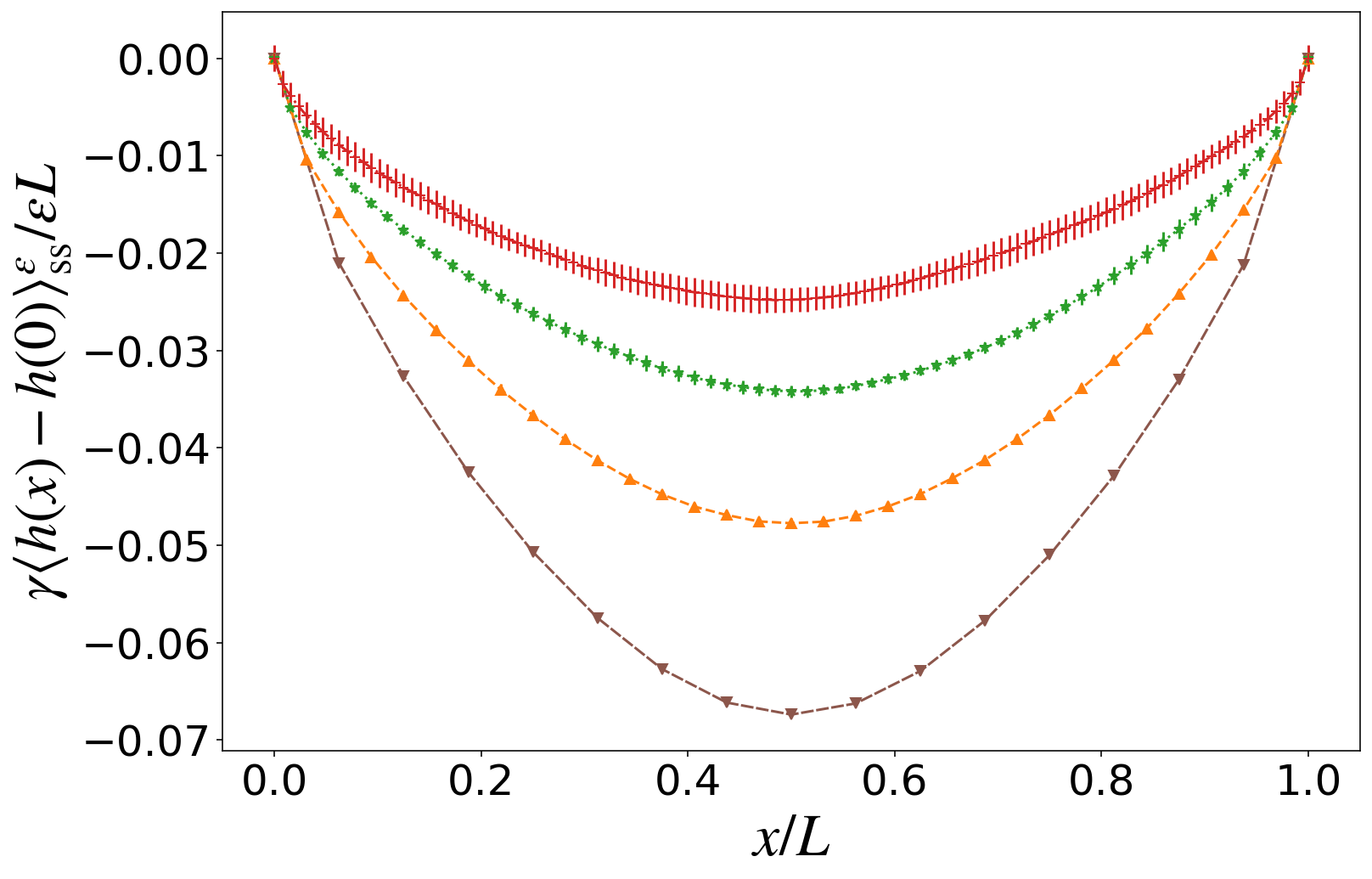}
    \caption{\label{fig:quad2} System size dependence of the curve for $v_0=5$ in Fig.\ref{fig:quad1}. The patterns of $L=8, 16, 32$, and $64$ are shown from
      {bottom to top}. {The {symbols} are joined by lines for visual aid.}
 Although the equilibrium ($v_0=0$) interface does not depend on $L$, the growing interface becomes stiffer for larger $L$.}
    \end{figure}


Now, let us consider a growing interface described by
\red{
\begin{equation}
\partial_t h = v_0+\frac{v_0}{2}(\partial_x h)^2
  -\frac{1}{\gamma}\var{F^\ep(\hat h)}{h} +\sqrt{\frac{2T}{\gamma}}\xi,
\label{ne-int}
\end{equation}
}
as a generalization of (\ref{eq-int}), where $v_0 \ge 0$ is
the propagation velocity of the flat interface.
When $\ep=0$, (\ref{ne-int}) is equivalent to the Kardar-Parisi-Zhang (KPZ) equation \cite{KPZ}, which {qualitatively} reproduces the dynamics of growing interfaces, such as {interfaces in liquid-crystal turbulence \cite{Takeuchi2012}, slow-combustion fronts in paper \cite{Maunuksela1997}, and fronts of growing bacterial colony \cite{Wakita1997}}. Because interfaces appear at almost all scales of interest in science \cite{KPZ_review_T2018, KPZ_review_HT2015}, the KPZ equation has been extensively investigated through numerical \cite{Meakin1993, Newman1996, Lam1998, Miranda2008, Wio2010, Dentz2016, Priyanka2020}, 
theoretical \cite{Beijeren1985, Medina1989_RG,  Frey1994_RG, Colaiori2001, 
Prahofer2004, Canet2011_RG, Niggemann2022}, and even 
mathematical \cite{Sasamoto2010, Dotsenko2010, Calabrese2010, Amir2011, 
Kloss2012, Imamura2012, Hairer2013, Johansson_2019} approaches. 
The system given by  (\ref{ne-int}) is interpreted as a perturbed
KPZ equation.

{\it Numerical observation.---}
Let $\bra h(x)-h(0) \ket_{\ss}^\ep$
be the expectation of $h(x)-h(0)$ with respect
to the steady state of (\ref{ne-int}). As an illustration, first,
we numerically investigate $\bra h(x)-h(0) \ket_{\ss}^\ep$ for 
the specific parameter 
values $\kappa=T=\gamma=1$ and $v_0=5$. Throughout this study, the 
numerical
simulations were conducted using a spatially discretized model
with a space interval $\Delta x=0.5$ \cite{SM, Lam1998}.
{More precisely, we define a discrete model and check system size
dependence to judge whether it gives a systematic  approximation of the
KPZ equation.}
The shapes of the growing interfaces shown in Fig. \ref{fig:quad2}
are fitted to the following form:
\begin{equation}
  \bra h(x)-h(0) \ket_{\ss}^\ep =\frac{\epsilon}{2 L \kappa_{\eff}}
  \left[ \left(x-\frac{L}{2} \right)^2-\frac{L^2}{4} \right],
\label{nss-shape}
\end{equation}
which is the generalization of (\ref{eq-shape}) with the replacement
of $\kappa$ by $\kappa_{\eff}$, where $\ep$ is assumed to be sufficiently 
small. The fitting parameter $\kappa_{\eff}$ 
is interpreted as the effective surface tension characterizing the
stiffness of the growing interface. 
We conjecture that (\ref{nss-shape}) is valid in the limit $\ep \to 0$ 
because the linear response for the noiseless case is expressed as a quadratic function \cite{SM}.
Fig. \ref{fig:quad1} 
shows that $\kappa_\eff$ is greater than $\kappa$. 
Furthermore, as shown in Fig. \ref{fig:quad2},
$\kappa_\eff$ increases for a larger system size $L$.

Now, two issues naturally arise. The first issue is quantifying the $L$ 
dependence of $\kappa_\eff$. From the viewpoint of numerical calculation, however, it becomes harder to accurately observe a slight shift of $\left<h(x)-h(0)\right>_{\rm ss}^\ep$ caused by the external stress for larger systems. The second issue is to investigate the mechanism of the $L$ dependence. Both issues can be resolved by formulating a fluctuation-response relation for the system under investigation, where $\kappa_\eff$ is expressed by dynamical properties of fluctuations in a system without the external stress.

{\it Response formula.---}
Let $[\hat h]$ be a trajectory $(\hat h(t))_{t=0}^\tau$.
We consider any quantity $A([\hat h])$ satisfying
$A([\hat h+\hat c])=A([\hat h])$, where $\hat c$ is a constant function in 
$x$.
For such $A([\hat h])$, we define $A^*$ as  $A^*([\hat h])\equiv A([\hat h]^*)$
with $[\hat h]^* \equiv (\hat h(\tau-t))_{t=0}^\tau$, which represents
the time-reversal of $[\hat h]$. For equilibrium cases $v_0=0$,
the detailed balance condition $\bra A \ket_\eq^\ep=\bra A^* \ket_\eq^\ep$
holds for any $\epsilon$, and the stationary distribution is given by
\begin{equation}
P^{\ep}_\eq(h)= \frac{1}{Z} \exp(-\beta F^{\ep}(h)) 
\end{equation}
with $\beta=T^{-1}$. This also leads to (\ref{eq-res}).

    \begin{figure}[t]
    \includegraphics[width=8cm]{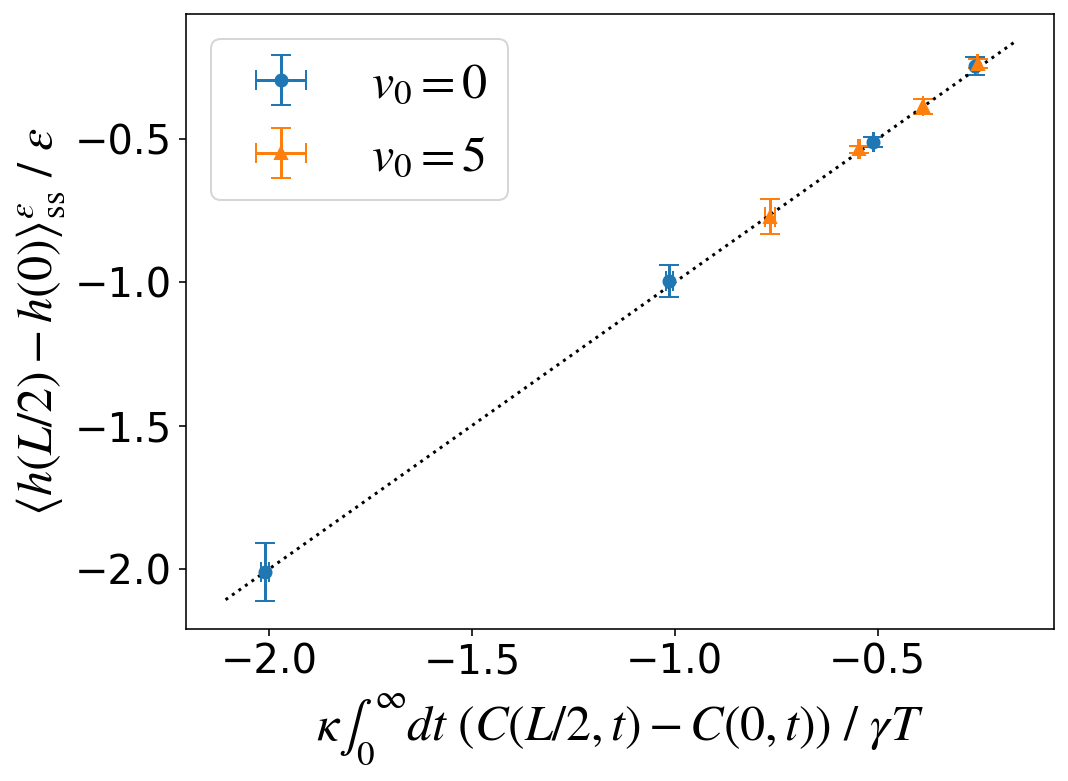}
    \caption{\label{fig: L h/ep}
      Comparison of the left-hand side and the right-hand side of \eqref{FRR}. 
      The former is estimated by the direct calculation of the response 
for $\ep>0$, while the latter is calculated in the system with $\ep=0$. 
The round-blue and triangular-orange symbols represent the data for 
$v_0=0$ ($L = 2, 4, 8, 16$) and $v_0=5$ ($L = 2, 4, 8, 16$), respectively. 
These symbols should be on the dotted line if the left and right-hand 
sides of \eqref{FRR}
      are equal.}
    \end{figure}

    \begin{figure}[t]
    \includegraphics[width=8cm]{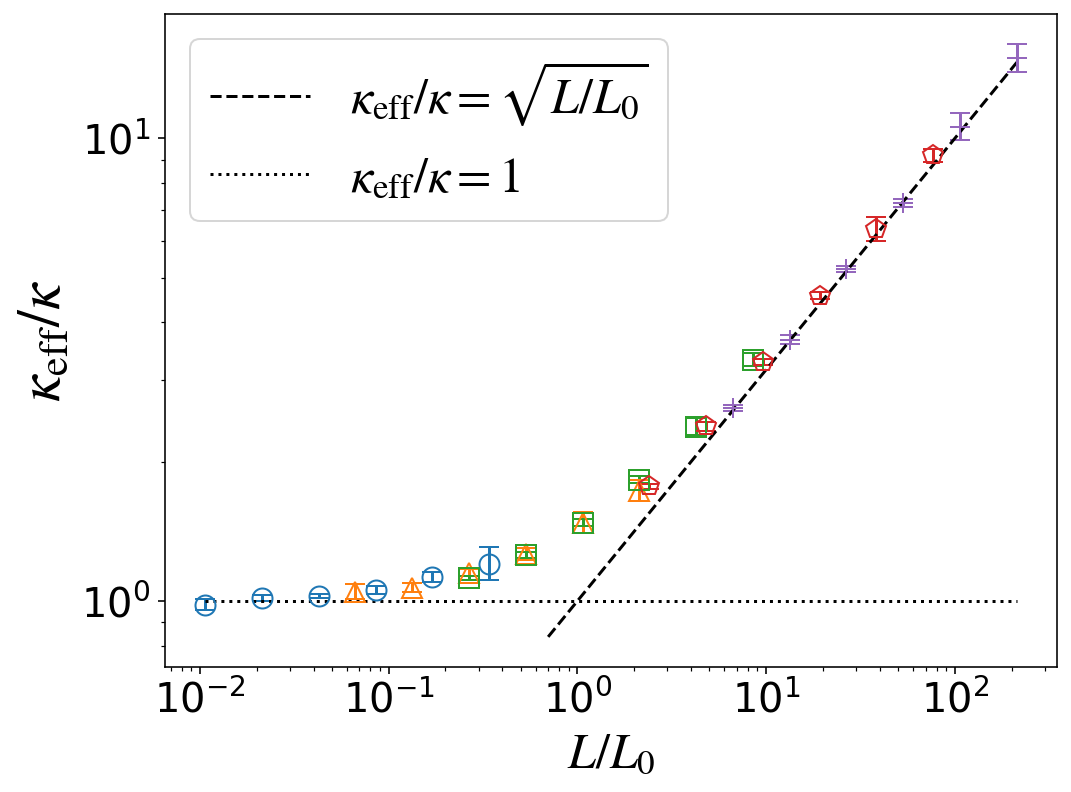}
    \caption{\label{fig: L nu}
      System size dependence of $\kappa_{\rm eff}$. The symbols are the numerical results for $L=16,32,64,128,256$, and $512$
      for $v_0=0.2, 0.5, 1,3$, and $5$ \red{from left to right (difference in symbols represents the difference in $v_0$)}.
      $\kappa_{\rm eff}$ maintains the same value as $\kappa$ for $L\ll L_0$ and diverges
      as $(L/L_0)^{1/2}$ for $L\gg L_0$.}
    \end{figure}

For growing interfaces with $v_0 >0$, the detailed balance condition does not hold. The extent of the violation is expressed by the entropy production 
\begin{equation}
  \sigma=\frac{\gamma}{T}\int_0^\tau dt\int_0^L dx\
  (\partial_t h) \left( v_0+\frac{v_0}{2}(\partial_x h)^2\right),
\label{sigma}
\end{equation}
which is the work done by the non-conservative force divided by
the temperature.
Using this thermodynamic entropy production, we
arrive at the standard fluctuation theorem \red{\cite{SM}}
\begin{equation}
    \bra A \ket_{\rm tr}^I=\bra A^* e^{-\sigma}\ket_{\tr}^I ,
\label{FT-1}
\end{equation}
where $\bra \cdot\ket_{\rm tr}^I$ denotes the ensemble average
over noise realizations and the initial conditions sampled from the
stationary distribution with $v_0=0$. \red{This relation holds for a wide range of driven systems in contact with a heat bath \cite{Evans1993, 
Gallavotti1995, Kurchan,SeifertRPP}.
However, \eqref{FT-1}
is not useful to obtain the linear response property around the
state with $v_0 \not =0$ and $\epsilon=0$. }

\red{Here, we notice }
another time-reversal transformation 
{$[\hat h]\to\ $}$[\hat h]^\dagger \equiv (-\hat h(\tau-t))_{t=0}^\tau$
such  that $\bra A \ket_\ss^0=\bra A^\dagger \ket_\ss^0$
holds for $A^\dagger([\hat h])\equiv A([\hat h]^\dagger)$ {\cite{FNS,SM}}.
However, this time-reversal symmetry is violated
for interfaces under the external stress $\epsilon >0$.
Then, following the standard procedure for the fluctuation
theorem \cite{SeifertRPP},
we calculate the ratio of path probabilities of $[\hat h]$
and $[\hat h]^\dagger$ and take the logarithm of the result to obtain
\begin{equation}
  \tilde \sigma \equiv -\frac{\epsilon \kappa}{\gamma T}
    \int_0^\tau dt\int_0^L dx\  p_\ex(x)\frac{\partial^2 
h(x,t)}{\partial^2 x},
\end{equation}
which characterizes the violation of the symmetry associated with
the time-reversal transformation $[\hat h] \to [\hat h]^\dagger$.
Indeed, we
can show a generalized fluctuation theorem
\begin{equation}
    \bra A \ket_{\rm tr}^{II}=\bra A^\dagger e^{-\tilde 
\sigma}\ket_{\tr}^{II}, 
\label{FT-2}
\end{equation}
where $\bra \cdot\ket_{\rm tr}^{II}$ denotes the ensemble average
over the noise realizations and initial conditions sampled from the 
stationary distribution without the external stress. 
Note that $\tilde \sigma$ is not the thermodynamic entropy production,
but interpreted as {an excess entropy production that appears
only when the external stress is imposed \cite{Sasa2000}.}

Here, we set $A=h(x,\tau)-h(0,\tau)$, substitute it into (\ref{FT-2}), take the limit $\tau \to \infty$, and expand the right-hand side of (\ref{FT-2}) in $\epsilon$. 
Noting that $\left< A \right>^{II}_{\rm tr}$ goes to
$\left< h(x)-h(0)\right>_{\rm ss}^{\ep}$, we obtain \cite{SM}
\begin{equation}
\lim_{\ep \to 0}\frac{\left<h(x)-h(0)\right>_{\rm 
ss}^{\epsilon}}{\epsilon}
= \frac{\kappa}{\gamma T}\int_0^\infty dt\ \left(C(x,t)-C(0,t)\right)
\label{FRR}
\end{equation}
with
\begin{equation}
  C(x,t)\equiv
  \left\langle\partial_x h(x,0) \partial_x h(0,t)\right\rangle_{\rm 
ss}^{0}.
\end{equation}
This relation is interpreted as the fluctuation-response relation 
of the system under investigation. (\ref{FRR}) is understood from the fluctuation-dissipation theorem for classical stochastic processes \cite{MSR,FDT,Jans}. However, to our best knowledge, an explicit formula connecting the response to an external perturbation has never been proposed to date. We numerically check the validity of (\ref{FRR}) for small systems with $L=2,4,8$, and $16$.
In Fig. \ref{fig: L h/ep},
the left-hand side of (\ref{FRR}) is plotted against
the right-hand side of (\ref{FRR}) at $x=L/2$ for both cases of $v_0=0$ and
$v_0=5$. The result confirms that (\ref{FRR}) holds. 


{\it Divergent stiffness.---}
As explained above, the numerical calculation of $\kappa_{\eff}$ 
defined by \eqref{nss-shape} is not easy to carry out for large systems. Thus, using the response formula (\ref{FRR}), we study the stiffness of the growing interface. Specifically, from  \eqref{nss-shape} and \eqref{FRR}, we obtain
\begin{equation}
 \kappa_{\rm eff}=-\frac{\gamma TL}{8\kappa}\left\{\int_0^\infty dt\
 \left[ C\left(\frac{L}{2},t\right)-C(0,t)\right]\right\}^{-1}.
 \label{Eq: nu_eff}
\end{equation}
By dimensional analysis, we find that $\kappa_{\eff}/\kappa$ is
expressed as a function of $L/L_0$ with
\begin{equation}
L_0= \frac{\ell \kappa^3}{T \gamma^2 v_0^2},
\label{L0}
\end{equation}  
where $\ell$ is a numerical constant corresponding to the dimensionless
length characterizing the cross-over \cite{SM}. 
In other words, the following equation is obtained using a scaling 
function $f$ whose form has not been determined yet:
\begin{equation}
  \kappa_{\rm eff}=\kappa f\left( \frac{L}{L_0} \right).
\end{equation}
First, we notice that $\kappa_\eff \to \kappa$ as $L_0 \to \infty$,
because $v_0 \to 0$ refers to the equilibrium limit. To find the 
functional
form of $f$, the right-hand side of (\ref{Eq: nu_eff}) is numerically calculated for several values of $L$ 
and $v_0$ for fixed
$\kappa=T=\gamma=1$. The numerical results are plotted in 
Fig. \ref{fig: L nu}, such that the following equation holds for $L \gg 
L_0$:
\begin{equation}
 \kappa_{\rm eff}=\kappa \left( \frac{L}{L_0} \right)^{1/2},
\end{equation} 
as indicated by the dotted line in Fig. \ref{fig: L nu}. 
\red{Here, the value of $\ell$ is numerically estimated as $\ell=60$.} It is found that the data points for $L \ge 16$ are on one curve, which determines the form of the scaling function $f$. Note that those for $L \le 8$\red{, which are not shown in Fig. \ref{fig: L nu},} deviate from
the curve \cite{SM}. \red{This means that the discretized equation used for the numerical calculation is no longer} 
a good approximation of the KPZ equation \red{when $L \le 8$}.
From Fig. \ref{fig: L nu}, it is concluded that $L_0$ with $\ell\simeq 60$ provides
the cross-over length from the normal response to the singular response,
where the stiffness $\kappa_{\rm eff}$ shows the divergence as a function
of $L/L_0$,  which is the main result of this Letter.


The divergent stiffness comes from a dynamical singularity of the correlation function $C(x,t)$, as suggested in the formula expressed by  \eqref{Eq: nu_eff}. The relation is explained in detail. Let $\tilde C(k,t)$ be the Fourier transform of $C(x,t)$. By dimensional analysis, we have
    \begin{equation}
      \int_0^\infty dt \ \tilde C(k,t)=
      \frac{\gamma T}{\kappa^2 k^2} \Phi\left(
      \frac{k \kappa^3}{T \gamma^2 v_0^2} \right),
\label{c-int}
    \end{equation}
    {where the prefactor is the equilibrium form and
      the non-equilibrium correction is expressed in terms of
      a dimensionless scaling function $\Phi$.} 
    Now, let us consider the case $L \to \infty$ with fixed $v_0 \not 
=0$.
    As is known,
    $\tilde C(k,t)$ has the scaling form $g(k^z t)$ in the limit $L \to 
\infty$,
    where the dynamical exponent $z$ is given by  $z=3/2$ \cite{FNS,KPZ}.
    {Assuming that the scaling part of $\tilde C(k,t)$ is dominant
      for the evaluation of $\kappa_{\rm eff}$, we 
  substitute the scaling form  into the left-hand side of
  (\ref{c-int}).  We then obtain} \cite{SM}
\begin{equation}    
\Phi\left(  \frac{k \kappa^3}{T \gamma^2 v_0^2} \right)
=  c \left(\frac{|k| \kappa^3}{T \gamma^2 v_0^2} \right)^{1/2},
\label{phi-asym}
\end{equation}
where the numerical constant $c$ is calculated as $c=2.43$
by the analysis of an exactly solvable stochastic model \cite{Prahofer2004}. For finite
but large $L$ cases, it is assumed that (\ref{phi-asym})
holds with the replacement of $k$ by $k_n =2\pi n /L$, where $n$ is
an integer satisfying $-\nc \le n \le \nc$.  The cutoff integer
$\nc$ is given by $\nc=L/(2 \Delta x)$. 
We then {calculate} \cite{SM}
\begin{align}
&  \int_0^\infty dt\ [C(L/2,t)-C(0,t)] \nonumber \\
&  = - \sqrt{L} \left( \frac{16 c^2}{8 \pi^3}  \right)^{1/2}
  \left(\frac{T}{\kappa v_0^2} \right)^{1/2}
\sum_{n=1}^{\nc/2} \frac{1}{(2n-1)^{3/2}}.
\label{c-est}
\end{align}
By substituting \eqref{c-est} into \eqref{Eq: nu_eff}, 
$\kappa_{\rm eff}=\kappa (L/L_0^{\est})^{1/2} $
holds with $L_0^{\rm est}= \ell^{\est} \kappa^3/(T \gamma^2 v_0^2)$,
where the numerical constant $\ell^{\rm est}$ is {given} as
\begin{equation}
\ell^{\est}= \frac{(32c)^2}{(2\pi)^3}
\left( \sum_{n=1}^{\nc/2} \frac{1}{(2n-1)^{3/2}} \right)^2.
\label{ell-est}
\end{equation}
Therefore, the divergent stiffness arises from the singularity
expressed by \eqref{phi-asym}. The $\sqrt{L}$ dependence of $\kappa_{\rm 
eff}$
corresponds to the $k^{3/2}$ dependence of
$ \int_0^\infty dt\ \tilde C(k,t)$. 
The cross-over length $L_0$ observed in the numerical
simulations is predictable by considering the asymptotic form of $ \int_0^\infty dt\ \tilde C(k,t)$. Indeed, the value $\ell\simeq 60$ obtained by the numerical simulations is consistent with (\ref{ell-est}). For example, $\ell^{\est}=62.5$
for $\nc=128$. When investigating infinitely large systems, the limit $n_c\to\infty$ should be taken. In this case, $\ell^{\est}$ approaches $69.52$ \cite{SM}.

{\it Concluding remarks.---}    
In this Letter, the response formula \eqref{Eq: nu_eff} expressing the effective surface tension is formulated in terms of the time correlation function $C(x,t)$ of $\partial_x h(x,t)$. Then, it is shown that the divergent stiffness comes from the dynamical singularity expressed by \eqref{phi-asym}.
    
    
The stochastic dynamics of the interface can be observed in a 
much wider context \cite{Takeuchi2014}. Keeping the universality in mind, we study the KPZ equation 
\begin{equation}
  \partial_t h=\frac{\lambda}{2}(\partial_x h)^2+ \nu
  \partial_x^2h  + \sqrt{2D}\xi 
   \label{Eq: KPZ}
\end{equation}
defined in $0 \le x \le L$, \red{where the standard parameters $\nu$, $D$, and $\lambda$ are introduced 
}.
By adding a localized 
force, $\nu_{\eff}$ instead of $\kappa_{\eff}$ can be operationally defined through (\ref{nss-shape}). 
Our formula \red{\eqref{L0} with replacements $\kappa/\gamma\to\nu$, $v_0\to\lambda$, and $T/\gamma\to D$} can be used to estimate $\nu$, $D$, and $\lambda$ 
when there exists a phenomenon that may be effectively described by
the KPZ equation. Specifically, 
one can estimate $\nu^3/(D \lambda^2)$ by observing cross-over length of $\nu_{\rm eff}$. 
From the fluctuation spectrum of $\partial_x h$, $D/\nu$ is determined. The parameter $\lambda$ is determined from the average propagation velocity. These three data lead to $\nu$, $D$, and $\lambda$.
{For example, putting oil on boundaries of an interface in combustion of paper \cite{Maunuksela1997}, we can study a response property. Since the system is described by the model in this Letter, the parameter values of the KPZ equation
will be determined by using the method above. 


  We thank K. Takeuchi for fruitful discussions.
  This study was supported by JSPS KAKENHI (Grant Numbers JP19H05795,
  JP20K20425, and JP22H01144). 




  

\onecolumngrid
\newpage


\newgeometry{margin=25truemm}

{\centerline{\large\textbf{Supplemental Material:}}}
\vspace{1mm}
{\centerline{\large\textbf{Divergent stiffness of a growing interface}}}
\vspace{5mm}
\centerline{Mutsumi Minoguchi and Shin-ichi Sasa}
\vspace{10mm}

\setcounter{equation}{0}
\renewcommand{\theequation}{S.\arabic{equation}}
\setcounter{figure}{0}
\renewcommand{\thefigure}{S.\arabic{figure}}

Throughout the supplemental material, we set $\nu=\kappa/\gamma$,
$D=T/\gamma$, $\lambda=v_0$, and $\tep=\ep/\gamma$. The equation
we study is then expressed as
\begin{equation}
  \partial_t h
  =\nu\partial_x^2h+\frac{\lambda}{2}(\partial_x h)^2
+\sqrt{2D}\xi + \tep\ p_\ex(x), 
\label{Eq: KPZ+F}
\end{equation}
which is a forced KPZ equation with the most standard notation. The noise $\xi=\xi(x,t)$ satisfies
\begin{align}
\left<\xi(x,t)\right>&=0, \\
\left<\xi(x,t)\xi(y,s)\right>
&=\delta(x-y)\delta(t-s).
\end{align}
We particularly focus on the case
\begin{equation}
  p_\ex(x)=\delta(x)-\frac{1}{L},
\label{pex}  
\end{equation}  
and the periodic boundary condition is assumed for
$0 \le x \le L$.

The organization of the supplemental material is as follows.
First, we summarize the basic issues of the equation we study.
Then, in Sec. \ref{derivation}, we derive the formulas
presented in the main text. In Sec. \ref{sm-data},
we show some numerical results supporting arguments in
the main text.

\section{Basic issues}\label{basic}

\subsection{Dimensional analysis}

A solution of (\ref{Eq: KPZ+F}) with a parameter set
$(\nu, D, \lambda, L, \tep)$ is connected to another solution
of (\ref{Eq: KPZ+F}) with a different parameter set
$(\nu', D', \lambda', L', \tep')$ by some scaling transformations.
We explicitly confirm this fact, which is useful to derive a
general expression of the effective surface tension. 

First, we introduce coordinate transformations
\begin{subequations}
\begin{align}
    x &=\alpha_x x' ,\\
    t &=\alpha_t t' ,
\end{align}
\label{Eq: transform}
\end{subequations}
where $\alpha_x >0$ and $\alpha_t >0$ are scaling factors.
Accordingly, $0 \le x' \le L'$ with $L'=L/\alpha_x$.  We also define
$h'(x',t')$ by 
\begin{align}
h(x,t) &=\alpha_h h'(x',t')
\end{align}
with a new scaling factor $\alpha_h >0$. Finally, we introduce
$\xi'(x',t')$ that satisfies
\begin{align}
\left<\xi'(x',t')\right>&=0,\\
\left<\xi'(x',t')\xi'(y',s')\right>&=\delta(x'-y')\delta(t'-s').
\end{align}
This is explicitly written as
\begin{align}
\xi(x,t) &=\alpha_x^{1/2}\alpha_t^{1/2} \xi'(x',t').
\end{align}
By substituting these relations into \eqref{Eq: KPZ+F} with \eqref{pex},
we obtain 
\begin{equation}
   \partial_{t'} h'=\alpha_t\alpha_x^{-2}\nu\partial_{x'}^2h'
   +\alpha_t\alpha_x^{-2}\alpha_h\frac{\lambda}{2}(\partial_{x'} h')^2
   +\alpha_x^{-1/2}\alpha_t^{1/2}\alpha_h^{-1}\sqrt{2D}\xi' 
   + \alpha_t\alpha_x^{-1}\alpha_h^{-1} \tep\
   \left(\delta(x')-\frac{1}{L'} \right).
\label{Eq: KPZ+F-2}
\end{equation}
From this expression, we define a new set of parameters
$(\nu', D', \lambda', \tep') $ by 
\begin{subequations}
\begin{align}
    \alpha_t\alpha_x^{-2}\nu &=\nu' , \\
    \alpha_t\alpha_x^{-2}\alpha_h\frac{\lambda}{2} &=\frac{\lambda'}{2},\\
    \alpha_x^{-1/2}\alpha_t^{1/2}\alpha_h^{-1}\sqrt{2D} &= \sqrt{2D'}, \\
    \alpha_t\alpha_x^{-1}\alpha_h^{-1} \tep&=\tep',
\end{align}
\label{Eq: param transform}
\end{subequations}
so that \eqref{Eq: KPZ+F-2} becomes 
\begin{equation}
  \partial_{t'} h'=\nu'\partial_{x'}^2h'
  +\frac{\lambda'}{2}(\partial_{x'} h')^2
  +\sqrt{2D'}\xi' + \tep'\ \left(\delta(x')-\frac{1}{L'} \right).
   \label{Eq: KPZ+Fd}
\end{equation}
By solving \eqref{Eq: param transform}, we find 
\begin{subequations}
\begin{align}
    \alpha_x &=\frac{r_\nu^3}{r_D r_\lambda^2},\\
    \alpha_t &=\frac{r_\nu^5}{r_D^2 r_\lambda^4},\\
    \alpha_h &=\frac{r_\nu}{r_\lambda},\\
    r_{\tep}&=\frac{r_\lambda r_D}{r_\nu},
\end{align}
\end{subequations}
where $r_\nu=\nu/\nu'$, $r_\lambda=\lambda/\lambda'$, $r_D=D/D'$, and $r_{\tep}=\tep/\tep'$. 
That is, by appropriately choosing $\alpha_x$, $\alpha_t$, and $\alpha_h$ for 
\eqref{Eq: KPZ+F}, we obtain a different model with $\nu'$, $D'$, and $\lambda'$ whose values are specified.

\subsection{Discrete model}\label{Dis:model}

In numerical simulations of \eqref{Eq: KPZ+F}, we define a discrete field $h_i(t)\equiv h(i\Delta x,t)$ with $0 \le i \le N$, where $\Delta x$ is a numerical parameter,  $N \Delta x=L$, and $h_0=h_N$. Throughout this study, we fix $\Delta x=0.5$.

For later convenience, we also define \begin{equation} 
u_i(t) \equiv \frac{h_{i+1}(t)-h_i(t)}{\Delta x},
\end{equation} 
which represents a discrete field corresponding to $u(x,t) = \partial_x h(x,t)$.
{The rule of discretization is that $h_i$ is defined on $i$-site
and $u_i$ is defined on the bond connecting $i+1$-site and $i$-{s}ite.
This clearly represents the conservation law and the symmetry which
is described below {in this and the next sections}.}
We then study the following discrete model of \eqref{Eq: KPZ+F} \red{\cite{Lam1998}}:
\begin{equation}
    \frac{d h_i}{dt}
    =\nu \frac{u_{i}-u_{i-1}}{\Delta x}
    + \frac{\lambda}{6}\left(u_i^2+u_iu_{i-1}+u_{i-1}^2\right)
    +\tep\ p_{\ex}{}_i
    + \sqrt{\frac{2D}{\Delta x}}\xi_i,
\label{KPZ-lan}
\end{equation}
where $\xi_i(t)$ is  Gaussian white noise that satisfies
$\left<\xi_i(t)\right>=0$ and $\left<\xi_i(t)\xi_j(t')\right>
=\delta_{ij}\delta(t-t')$. $\tep\ p_{\ex}{}_i$ is a discrete form
of the external stress $\ep p_\ex(x)$. \red{Note that $\nu$, $\lambda$, and $D$ are the parameters introduced at the beginning of Supplemental Material.}
It can be seen that (\ref{KPZ-lan}) leads to \eqref{Eq: KPZ+F}
in the limit $\Delta x \to 0$.

We express $(u_0,\cdots, u_{N-1})$ as $\bv{u}$ collectively.
The stationary distribution of $\bv{u}$ for $\ep=0$ is
then calculated as 
\begin{equation}
  P^{\rm ss}_{0}(\bv{u})
  = \left(\frac{\nu (\Delta x)}{2 D \pi} \right)^{N/2}
  \exp\left[- \frac{\nu}{2D} \sum_{i}(\Delta x) u_i^2  \right].
\label{ss-lan}
\end{equation}
More precisely, the non-linear term in (\ref{KPZ-lan}) is
chosen so that (\ref{ss-lan}) holds \cite{Lam1998}.
See \eqref{key-dis} for the argument. 
We used  the Heun method to solve (\ref{KPZ-lan}) numerically,
and we confirmed that the numerical estimation of $\bra u_i^2 \ket$
is equal to the theoretical value within 1\% error for $\Delta t=0.01$.

Here, we note that  \eqref{KPZ-lan} is rewritten as the continuity equation 
\begin{equation}
    \frac{d  u_i}{dt}+\frac{j_{i+1}-j_i}{\Delta x}=0,
\end{equation}
where the current field is given as 
\begin{equation}
  j_i \equiv
  -\frac{\lambda}{6}\left(u_i^2+u_iu_{i-1}+u_{i-1}^2\right)
  - \nu \frac{u_{i}-u_{i-1}}{\Delta x}
  -\tep\ p_{\ex}{}_i
  - \sqrt{\frac{2D}{\Delta x}}\xi_i .
\end{equation}
The trajectory $(\bv{u}(t))_{t=0}^\tau$ is collectively denoted by $[\bv{u}]$.
Let ${\cal P}([\bv{u}]|\bv{u}(0))$ be the probability density of trajectory
$[\bv{u}]$ provided that $\bv{u}(0)$ is given. From the Gaussian nature
of $\xi_i$, we first have
\begin{equation}
  {\cal P}([\bv{u}]|\bv{u}(0)) = {\cal N} \exp\left[
  -\frac{\Delta x}{4D}\int_0^\tau dt\ \sum_{i=1}^{N}
  \left(
  \frac{\lambda}{6}\left(u_i^2+u_iu_{i-1}+u_{i-1}^2\right)
  +\nu \frac{u_{i}-u_{i-1}}{\Delta x}
  +\tep\ p_{\ex}{}_i
  +j_i
  \right)^2
  \right],
\label{path-p}  
\end{equation}  
where the periodic boundary conditions have been used
in the derivation. 
The time integration is evaluated as the mid-point discretization and ${\cal N}$ is the normalization constant which depends on the time interval $\Delta t$ in the time integration.

\subsection{Time-reversal symmetry}\label{Dis:path}

We define the following two types of time-reversal transformation
{for {any} trajectory $[\bv{u}]$:
$[\bv{u}]^* \equiv (\bv{u}(\tau-t))_{t=0}^\tau$ and
$[\bv{u}]^\dagger \equiv (-\bv{u}(\tau-t))_{t=0}^\tau$.}
Associated with them, we have the two relations respectively.
The first relation is  
\begin{equation}
    \frac{{\cal P}([\bv{u}]^*|\bv{u}(\tau))}
         {{\cal P}([\bv{u}]||\bv{u}(0))}
    =\exp(-\sigma([\bv{u}]))
     \exp(\beta F^\ep(\bv{h}(\tau)) -\beta F^\ep(\bv{h}(0)))
\label{LDB-1}
\end{equation}
with
\begin{equation}
\sigma([\bv{u}])\equiv   
\frac{1}{D}\int_0^\tau dt\  \sum_i (\Delta x) \der{h_i}{t} 
\frac{\lambda}{6}\left(u_i^2+u_iu_{i-1}+u_{i-1}^2\right)
\end{equation}
and
\begin{equation}
F^{\ep}(\bv{h})=\gamma \sum_i (\Delta x) \left[\frac{\nu}{2} u_i^2- \tilde\ep p_\ex{}_i  h_i \right].
\end{equation}
Here, in the calculation of (\ref{LDB-1}), we have used
\begin{equation}
\der{F^\ep(\bv{h}(t))}{t}
= \sum_i \der{h_i}{t}
\frac{\partial F^\ep(\bv{h})}{\partial h_i}.
\end{equation}
When $\lambda=0$, \eqref{LDB-1} implies the detailed balance
condition with respect to the equilibrium distribution
proportional to $\exp[-\beta F^\ep(\bv{h})]$.

The second relation is 
\begin{equation}
  \frac{{\cal P}([\bv{u}]^\dag|-\bv{u}(\tau))}
       {{\cal P}([\bv{u}]||\bv{u}(0))}
    =\exp(-\tilde \sigma([\bv{u}]) )
     \exp\left( \frac{\nu}{2D}\sum_i u_i^2(\tau)\Delta x 
               -\frac{\nu}{2D}\sum_i u_i^2(0)\Delta x   \right)
\label{LDB-2}
\end{equation}
with  
\begin{equation}
\tilde \sigma([\bv{u}])\equiv   
-\frac{\nu}{D}\int_0^\tau dt\  \sum_i\tep p_{\ex}{}_i (u_i-u_{i-1}).
\end{equation}
{We have derived (\ref{LDB-2}) by substituting $ [\bv{u}]^\dag$
into  $ [\bv{u}]$ in (\ref{path-p}) and calculating the left-hand
side {of (\ref{LDB-2})}. In this calculation,}  we have used
\begin{align}
    \sum_i\partial_t (u_i^2)
    &=-\frac{2}{\Delta x}\sum_i(u_ij_{i+1}-u_ij_i)\\
    &=\frac{2}{\Delta x}\sum_i(u_i-u_{i-1})j_i ,
\end{align}
and
\begin{equation}
        \sum_i\left(u_i^2+u_iu_{i-1}+u_{i-1}^2\right)(u_{i}-u_{i-1})=0.
\label{key-dis}
\end{equation}
When $\ep=0$, \eqref{LDB-2} implies the detailed balance condition
with respect to the distribution proportional to
$\exp[-\nu/(2D)\sum_i u_i^2\Delta x]$. This provides the
proof of (\ref{ss-lan}).  Note that
$\nu/(2D)\sum_i u_i^2\Delta x$ is equal to
$\beta F^{0}(\bv{h})$.

\section{Derivation of formulas}\label{derivation}

\subsection{Derivation of (5)}

For equilibrium cases \red{(\ref{Eq: KPZ+F}) with $\lambda=0$},
the expectation of $h(x)$
under the external stress is determined by
\begin{equation}
  \nu \frac{\partial^2 }{\partial x^2}
  \left< h \right>_{\eq}^\ep   +\tep\left( \delta(x)-\frac{1}{L} \right)=\frac{\partial }{\partial t}
  \left< h \right>_{\eq}^\ep=0. 
\label{eq-profile}
\end{equation}
Since \red{$\partial_x^2 \bra h(x) \ket_\eq^\ep=\tilde\ep/(L\nu)$} for $x \not =0$
and $ \bra h(x) \ket_\eq^\ep=\bra h(L-x) \ket_\eq^\ep$,
we have
\red{
\begin{equation}
\left< h(x) \right>_{\eq}^\ep = \frac{\tep}{2L\nu}
  \left[\left(x-\frac{L}{2}\right)^2-\frac{L^2}{4}\right] + \left< h(0) \right>_{\eq}^\ep.
\end{equation}
}
Thus, 
\begin{equation}
  \left<h(x)-h(0)\right>_\eq^\ep
  =\frac{\tep}{2L\nu}
  \left[\left(x-\frac{L}{2}\right)^2-\frac{L^2}{4}\right].
\end{equation}
By setting $x=L/2$, we also have 
\begin{equation}
    \left<h(L/2)-h(0)\right>_\eq^\ep=-\frac{L\tep}{8\nu}.
\end{equation}

\subsection{\red{Non-linear response}}

{
In the main text, we conjecture the linear response form (7) for small
$\epsilon$. We here discuss this form from the nonlinear response
form for the noiseless case. 
}

{
The steady-state profile of $u(x) = \partial_x h(x)$ under
the external stress is determined by
\begin{equation}
  0=\frac{\partial}{\partial x}\left[\nu\frac{\partial u}{\partial x}+\frac{\lambda}{2} u^2+
    \tilde \epsilon\left(\delta(x)-\frac{1}{L} \right)\right],
    \label{Eq: KPZ deterministic F>0}
\end{equation}
\red{which is obtained from (\ref{Eq: KPZ+F}) with $\partial_t u=0$.}
\red{Equation  (\ref{Eq: KPZ deterministic F>0})} yields 
\begin{equation}
  \frac{B}{L}
  =\nu\frac{\partial u}{\partial x}+\frac{\lambda}{2} u^2
  +\tilde \epsilon\left(\delta(x)-\frac{1}{L} \right)
    \label{Eq: KPZ deterministic F>0 integrated}
\end{equation}
with a constant $B$. Solving this equation for $x \not =0$, we have
\begin{equation}
  u(x)=u_0 \tanh\left(\frac{\lambda u_0}{2\nu}
  \left(x-\frac{L}{2}\right)\right)
  \label{Eq: tanh}
\end{equation}
with
\begin{equation}
  u_0^2=\frac{2(B+\tilde\epsilon)}{\lambda L}.
\label{u0}
\end{equation}
Next, integrating  (\ref{Eq: KPZ deterministic F>0 integrated})
in the region $[0,\delta]$ and $[L-\delta,L]$, and  taking the
limit $\delta \to 0$, we obtain
\begin{equation}
  \nu( u(0+)-u(L-))+\tilde\epsilon=0.
\label{nr-bc}  
\end{equation}  
Combining it with (\ref{Eq: tanh}), we have
\begin{equation}
2{\nu}u_0 \tanh\left(\frac{\lambda u_0}{2\nu}
  \left(-\frac{L}{2}\right)\right)+\tilde \epsilon=0,
\label{u0-bc}  
\end{equation}  
which determines $u_0$ and $B$ from (\ref{u0}).
The integration of  (\ref{Eq: tanh}) in $x$ gives
\begin{equation}
\red{ h(x,t)=
  \frac{2\nu}{\lambda}\log
  \left[\cosh\left(\frac{\lambda u_0}{2\nu}
    \left(x-\frac{L}{2}\right)\right)\right] + h\left(\frac{L}{2},t \right)}
\label{nonlinear}
\end{equation}
This is the non-linear response form for the noiseless case.
}

Now, \red{we discuss the relation between (\ref{nonlinear}) and (7)
  in the main text. In the linear response regime of the limit
$\tilde \epsilon \to 0$}, from (\ref{u0-bc}), we find 
\begin{equation}
  u_0^2=\frac{2 \tilde\epsilon}{\lambda L}+
  {\mathcal{O}(\ep^2)}.
\label{u-bc1}
\end{equation}
Recalling (\ref{u0}), we also have $B=O(\tilde \epsilon^2)$.
Therefore, the profile in the linear response regime is given by
\begin{equation}
  u(x)=\frac{\tilde \epsilon}{L\nu}
  \left(x-\frac{L}{2}\right)+{\mathcal{O}(\ep^2)},
\end{equation}
which {leads to} 
\begin{equation}
h(x)-h(0)=\frac{\tilde \epsilon}{2L\nu}\left[\left(x-\frac{L}{2}\right)^2
    - \frac{L^2}{4} \right]{+\mathcal{O}(\ep^2)}.
\end{equation}
{This} is equivalent to (5) in the {main} text. That is, even for
growing interfaces, the linear response form is expressed as the
right-hand side of (5) if noise effects are ignored. Since it is
reasonably expected that the parameter $\nu$ is renormalized as
$\nu_{\rm eff}$ by noise effects, we conjecture (7) for growing
interfaces.

\subsection{Derivation of (10) and (11)}

In this section, we study the discrete model defined in
Sec. \ref{Dis:model}. 
We first note that $\bv{h}$ is not uniquely determined from
$\bv{u}$, because an additive constant of $\bv{h}$ is arbitrary for given
$\bv{u}$. 
Nevertheless, if a quantity $A(\bv{h})$ satisfies $A(\bv{h}+c \bv{1})=A(\bv{h})$
for any $c$, where $\bv{1}=(1,1,\cdots,1)$, we can uniquely express $A$ in terms of $\bv{u}$.
We thus write $A([\bv{u}])$, which we study in this section.

We first define 
\begin{equation}
  \bra A \ket_{I}^{\rm tr}
  \equiv \int {\cal D}[\bv{u}]
  A([\bv{u}]) {\cal P}(\bv{u}|\bv{u}(0))
  P^{\ep}_{\rm eq}(\bv{u}(0)),
\end{equation}
which represents the ensemble average
of $A$ over noise realizations and initial conditions sampled
from the equilibrium distribution with $\lambda=0$. 
Using (\ref{LDB-1}), we obtain
\begin{align}
\bra A \ket_{I}^{\rm tr}
&=\int \mathcal{D}[\bv{u}]
A([\bv{u}]^*)
\frac{{\cal P}([\bv{u}]^*|\bv{u}(\tau))}
     {{\cal P}([\bv{u}]|\bv{u}(0))}
     {\cal P}([\bv{u}]|\bv{u}(0))
     P^{\ep}_{\rm eq}(\bv{u}(t=\tau))\\
&=
\int \mathcal{D}[\bv{u}]
A([\bv{u}]^*)\exp(-\sigma([\bv{u}]))
     {\cal P}([\bv{u}]|\bv{u}(0))
     P^{\ep}_{\rm eq}(\bv{u}(t=0))\\
&=     
  \bra A^* e^{-\sigma} \ket^{\rm tr}_I,
\label{FT-1}
\end{align}
where $A^*([\bv{u}])\equiv A([\bv{u}]^*)$.
This is the standard form of the fluctuation theorem,
and $\sigma$ corresponds to the thermodynamic entropy
production in the system we study.

Since we have the second relation (\ref{LDB-2})
associated with time-reversal transformation, 
we further define 
\begin{equation}
 \bra A \ket_{II}^{\rm tr} \equiv \int {\cal D}[\bv{u}]
  A([\bv{u}]) {\cal P}(\bv{u}|\bv{u}(0)) P^{0}_{\rm ss}(\bv{u}(0)),
\end{equation}
which represents the ensemble average
of $A$ over noise realizations and initial conditions sampled
from the stationary distribution with $\ep=0$. 
Using (\ref{LDB-2}), we obtain
\begin{align}
\bra A \ket_{II}^{\rm tr}
&=\int \mathcal{D}[\bv{u}]
A([\bv{u}]^\dagger)
\frac{{\cal P}([\bv{u}]^\dagger|-\bv{u}(\tau))}
     {{\cal P}([\bv{u}]|\bv{u}(0))}
     {\cal P}([\bv{u}]|\bv{u}(0))
     P^{0}_{\rm ss}(\bv{u}(t=\tau))\\
&=
\int \mathcal{D}[\bv{u}]
A([\bv{u}]^\dagger)\exp(-\tilde \sigma([\bv{u}]))
     {\cal P}([\bv{u}]|\bv{u}(0))
     P^{0}_{\rm ss}(\bv{u}(t=0))\\
&=     
  \bra A^\dagger e^{-\tilde \sigma} \ket^{\rm tr}_{II},
\label{FT-2}
\end{align}
where $A^\dagger([\bv{u}])\equiv A([\bv{u}]^\dagger)$.
This fluctuation theorem is not standard, since $\tilde \sigma$ does not correspond to the thermodynamic entropy. Instead, $\tilde \sigma$ is interpreted as the excess entropy production that characterizes the extent of the violation of time-reversal symmetry in the KPZ equation.

\subsection{Derivation of (13)}

Setting $A=u_i(\tau)$ and substituting $p_{\rm ex}{}_i
=\delta_{i0}/\Delta x-1/L$ into (\ref{FT-2}), we have
\begin{equation}
\bra u_i(\tau) \ket_{II}^{\tr}
=\bra -u_{i}(0)\ e^{\frac{\nu\tep}{D\Delta x}
  \int_0^\tau dt\ (u_{0,t}-u_{N-1,t})}
 \ket_{II}^{\tr} .
\end{equation}
Then we expand it in $\tep$ and take the limit $\tau \to \infty$.
Noting that
\begin{equation}
  \lim_{\tau \to \infty} \bra u_i(\tau) \ket_{II}^{\tr} =
  \bra u_i \ket_{\ss}^\ep,
\end{equation}
we obtain
\begin{align}
\bra u_i \ket_{\ss}^\ep
&=-\frac{\nu\tep}{D\Delta x}\int_0^\infty dt\
\bra u_{i}(0)(u_{0}(t)-u_{N-1}(t)) \ket_{\ss}^0  +\mathcal{O}(\ep^2)\\
&= \frac{\nu\tep}{D\Delta x}\int_0^\infty
dt\ \left(\left<u_{i+1}(0)u_{0}(t)\right>_{\ss}^0-\left<u_{i}(0)u_{0}(t)\right>_{\ss}^0
  \right)
+ \mathcal{O}(\ep^2).
\end{align}
Therefore, the response formula is derived as
\begin{align}
    \left<h_i-h_0\right>_{\ss}^\ep
    &=\sum_{j=0}^{i-1}\Delta x \left<u_j\right>_\ss^\ep \\
    &=\sum_{j=0}^{i-1}\frac{\nu\tep}{D}
    \int_0^\infty dt\ \left(\left<u_{j+1}(0)u_{0}(t)\right>_\ss^0
    -\left<u_{j}(0)u_{0}(t)\right>_\ss^0 \right)
    + \mathcal{O}(\ep^2)\\
    &=\frac{\nu\tep}{D}\int_0^\infty dt\
    \left(\left<u_{i}(0)u_{0}(t)\right>_\ss^0
    -\left<u_{0}(0)u_{0}(t)\right>_\ss^0\right)+ \mathcal{O}(\ep^2)\\
    &=\frac{\nu \tep}{D}\int_0^\infty dt\
    \left(C_{i}(t)-C_{0}(t)\right)
    + \mathcal{O}(\ep^2),
\end{align}
where $C_{i}(t)\equiv\left<u_{i}(0)u_{0}(t)\right>_\ss^0$.
Taking the limit $\Delta x \to 0$, we have
\begin{equation}
 \left<h(x)-h(0)\right>_{\ss}^\ep
=\frac{\nu \tep}{D}\int_0^\infty dt\
    \left(C(x,t)-C(0,t)\right)
    + \mathcal{O}(\ep^2)
\end{equation}
with $C(x,t)\equiv\left<u(x,0)u(0,t)\right>_\ss^0$.

\subsection{Derivation of (17)}

We consider $\nu_{\rm eff}/\nu$ for the model \eqref{Eq: KPZ+F}.
We choose $\alpha_x$, $\alpha_t$, and $\alpha_h$ such that
$\nu'=D'=\lambda'=1$. We then have $\nu_{\rm eff}'$ is given as
a function of $L'$, which is simply written as
\begin{equation}
\nu_{\rm eff}'=f\left( \frac{L'}{\ell} \right),
\end{equation}
where $\ell$ is a numerical constant characterizing the dimensionless crossover length. 
Since  $\nu_{\rm eff}/\nu=\nu_{\rm eff}'$, we obtain
\begin{equation}
\frac{\nu_{\rm eff}}{\nu}= f\left( \frac{L}{\ell\alpha_x} \right)
  =f \left( \frac{D \lambda^2 L}{\ell \nu^3} \right).
\end{equation}
Thus, setting
\begin{equation}
    L_0 = \ell \frac{\nu^3}{D \lambda^2},
\end{equation}
we obtain the formula (17) in the main text.

\subsection{Estimation of $c$ in (20)}

We define the space-time Fourier transform of $C(x,t)$ as
\begin{equation}
  \check C(k,\omega)\equiv \int_{-\infty}^\infty
  dx \int_{-\infty}^\infty dt\ e^{i(kx+\omega t)} C(x,t).
\end{equation}
We then have 
\begin{equation}
\int_{0}^\infty dt\ \tilde C(k,t)=\frac{1}{2} \check C(k,0),
\label{st1}
\end{equation}
because $\tilde C(k,t)=\tilde C(k,-t)$. Here, Pr\"{a}hofer and Spohn showed that
\begin{equation}
  \check C(k,0)=9.72216 k^{-3/2}
\label{ps-C}
\end{equation}
for an exactly solvable stochastic model that corresponds
to the KPZ equation with  $D/\nu=1$ and $\lambda=1/2$ \cite{Prahofer note}. From (19) and (20) in
the main text,
we write 
\begin{equation}
  \int_{0}^\infty dt\ \tilde C(k,t)=
  c \left( \frac{D}{\nu \lambda^2} \right)^{1/2} k^{-3/2}, 
\label{s51}
\end{equation}
which becomes
\begin{equation}
  \int_{0}^\infty dt\ \tilde C(k,t)=  2c k^{-3/2}
\end{equation}
for the system with $D/\nu=1$ and $\lambda=1/2$. 
Comparing this expression and (\ref{st1}) with (\ref{ps-C}),
we obtain $c=2.43054$.

\subsection{Derivation of (21)}

\begin{figure}[t]
      \begin{minipage}[t]{0.48\hsize}
        \center
        \includegraphics[width=7cm]{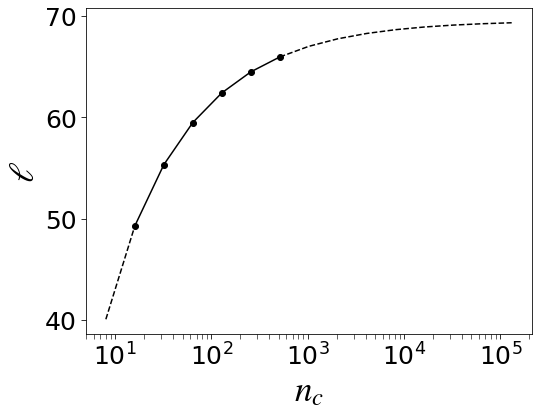}
        \caption{$\nc$ dependence of $\ell$
          expressed by \eqref{Eq: l}.
          The solid line shows the range of
         values we used in our numerical calculations.}
        \label{fig: l}
      \end{minipage}
      \hfill
      \begin{minipage}[t]{0.48\hsize}
        \center
        \includegraphics[width=7cm]{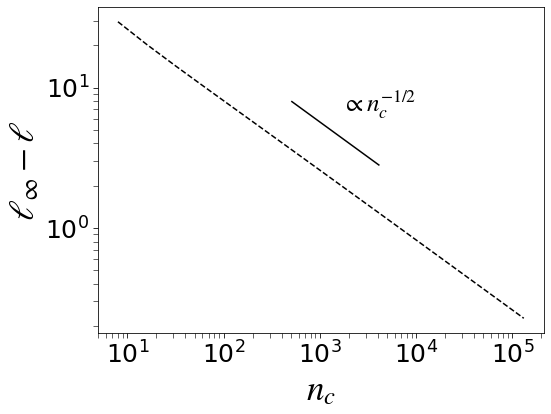}
        \caption{Asymptotic behavior of $\ell$ converging to $\ell_\infty$
          given in \eqref{Eq: linf}. The relation $\ell_\infty-\ell=80n_c^{-1/2}$ holds for large $n_c$.
          }
        \label{fig: l2}
      \end{minipage}
  \end{figure}

We first consider the Fourier expansion
\begin{equation}
    \int_0^\infty dt\ C(x,t)=\frac{1}{L}\sum_{n=-\nc}^{\nc} C_ne^{-ik_n x}
    \label{Eq: FT of C}
\end{equation}
with $k_n=2 \pi n/L$, where  the cut-off number $\nc$ is given by
$L/(2\Delta x)$.  From (\ref{s51}), we may set 
\begin{equation}
C_n=c \left( \frac{D}{\nu \lambda^2} \right)^{1/2} k_n^{-3/2}
\end{equation}
for sufficiently large $L$. We then find
\begin{align}
  &\int_0^\infty dt\ \left[ C\left(\frac{L}{2},t\right)-C(0,t)  \right]
  \nonumber \\
=&
\frac{c}{L} \left(\frac{D}{\nu \lambda^2} \right)^{1/2}
\sum_{n=-\nc}^{\nc}((-1)^{n}-1)\frac{1}{k_n^{3/2}} \nonumber \\
=&
-\frac{4c}{L} \left(\frac{L}{2\pi} \right)^{3/2}
\left(\frac{D}{\nu \lambda^2} \right)^{1/2}
\sum_{n=1}^{\nc/2}\frac{1}{(2n-1)^{3/2}}.
\end{align}
Substituting this result into the formula (15) in the main text,
we obtain  the form
\begin{equation}
\kappa_{\rm eff}= \kappa \left(\frac{L}{L_0}\right)^{1/2},
\end{equation}
where $L_0$ is calculated as 
\begin{equation}
  L_0=\ell\frac{\nu^3}{D \lambda^2}
  \label{Eq: L0}
\end{equation}
with
\begin{equation}
  \ell =\frac{(32c)^2}{(2\pi)^{3}}\left(\sum_{n=1}^{\nc/2}
  \frac{1}{(2n-1)^{3/2}}\right)^2.
  \label{Eq: l}
\end{equation}
This gives the estimation of the dimensionless cross-over length $\ell$
defined by (16) and (18) in the main text.

In Fig. \ref{fig: l}, we plot $\ell$ for  $\nc$.  In the range
we studied in numerical simulations, the theoretical estimation
of $\ell$ is consistent with the numerically obtained value $\ell\simeq 60$.
If we study the case that $L \to \infty$,  $\ell$ should be evaluated
in the limit $\nc\to\infty$. This value, which is denoted by $\ell_\infty$,
is calculated as
\begin{equation}
\begin{split}
\ell_\infty&=\frac{(32c)^2}{(2\pi)^{3}}\left(\sum_{n=1}^\infty \frac{1}{n^{\frac{3}{2}}}-\sum_{n=1}^\infty \frac{1}{(2n)^{\frac{3}{2}}}\right)^2\\
&=\frac{(32c)^2}{(2\pi)^{3}}\left(1-\frac{1}{2^{\frac{3}{2}}}\right)^2\zeta\left(\frac{3}{2}\right)^2\\
&\simeq 69.52.
\label{Eq: linf}
\end{split}
\end{equation}
Note that the convergence is slow and Fig. \ref{fig: l2} shows that 
\begin{equation}
\ell_\infty-\ell(\nc)= \frac{A}{\sqrt{\nc}}
\end{equation}
for large $\nc$, where $A=80$.

\section{Supplemental data} \label{sm-data}
    
\subsection{Measurement of the response}

\begin{figure}[t]
      \begin{minipage}[t]{0.48\hsize}
    \centering
    \includegraphics[width=7cm]{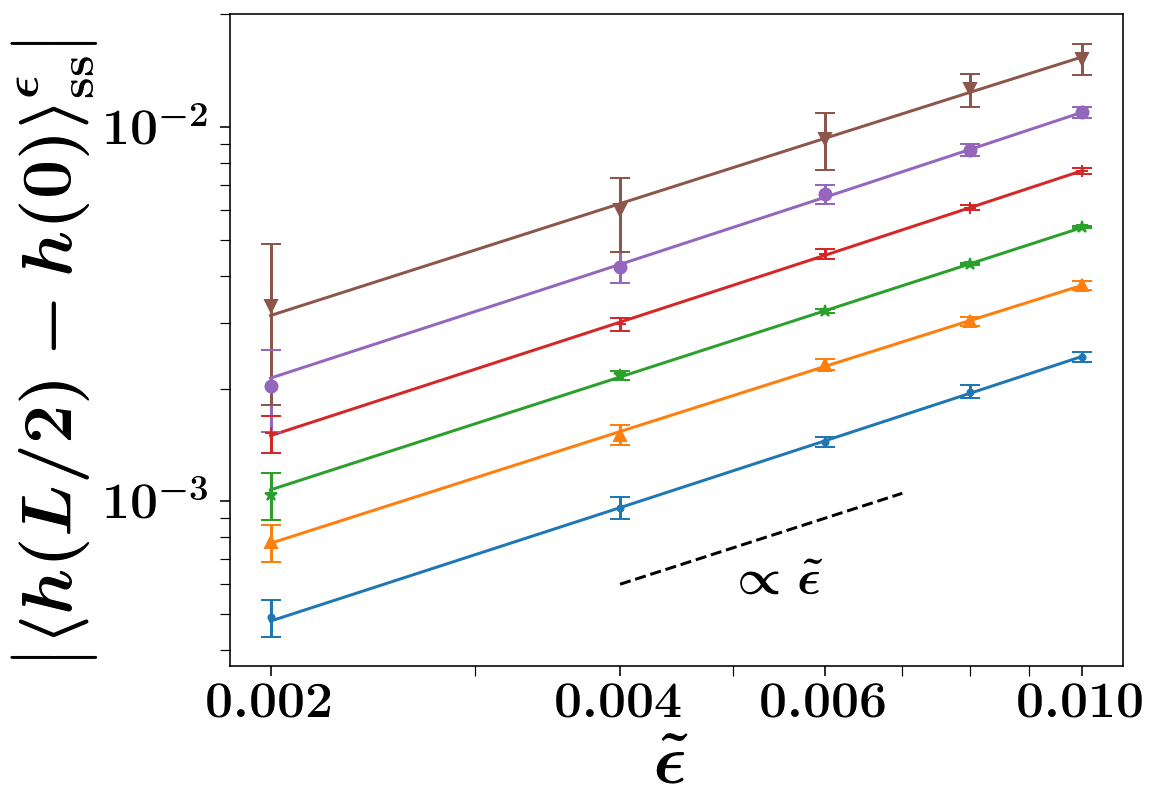}
    \caption{Amplitude of the averaged height function against the strength of the external stress for system sizes $L=2,4,8,16,32$, and $64$, from bottom to top. $v_0=5$. The symbols show the numerical results and they are fitted to a quadratic function $f_2=c_1\tilde\ep+c_2\tilde\ep^2$
       (solid lines). }
       \label{fig: ep_h}
      \end{minipage}
      \hfill
      \begin{minipage}[t]{0.48\hsize}
        \center
        \includegraphics[width=7cm]{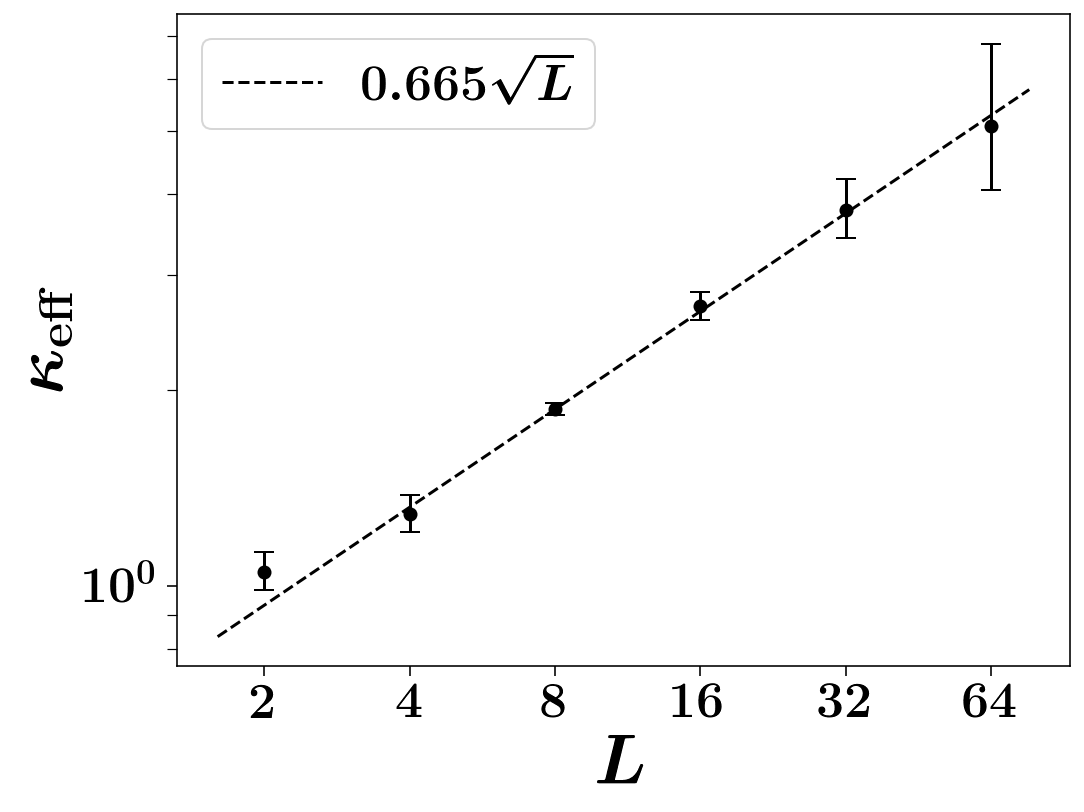}
        \caption{$\kappa_{\rm eff}$ calculated by \eqref{kappa-c1} are displayed as
          a function of $L$. $v_0=5$. This shows the divergent behavior
         $\kappa_{\rm eff} = 0.665\sqrt{L}.$}
        \label{fig: L_h}
      \end{minipage}
  \end{figure}

In Fig. 3 of the main text, we display the numerical data of $\left<h(L/2)-h(0)\right>_\ss^\ep/\tilde\ep$.
Here, we explain the method for numerically evaluating it. In Fig. \ref{fig: ep_h}, we show $\left<h(0)-h(L/2)\right>_{\ss}^\ep$ for several
values of $\tilde\ep$ ranged from $0.002$ to $0.01$ for
systems with sizes $L=2,4,8,16,32$, and $64$. For each $L$,
we fit the data points by a quadratic function $f_2(\tilde\ep)=c_1\tilde\ep+c_2\tilde\ep^2$
by the least-square method. This linear coefficient $c_1$ is given as $\bra h(L/2)-h(0)\ket^\epsilon_\ss/\tilde\epsilon$ in Fig. 3 of the main text.

Using the coefficient $c_1$ and (7) in the main text, we have
\begin{equation}
  \kappa_{\eff}=\frac{L}{8 c_1}.
  \label{kappa-c1}
\end{equation}  
Then, we display $\kappa_{\eff}$ as a function of $L$ in Fig. \ref{fig: L_h}.
This graph already shows the $\sqrt{L}$ behavior of $\kappa_{\eff}$.
However, it is hard to study larger systems than $L=64$ by this method.

\subsection{Dynamical-scaling exponent}

It has been known that the dynamical exponent $z$ is equal to $3/2$
for the KPZ equation. We directly confirm this fact by numerical simulations. Concretely, we study the
height width
\begin{equation}
    W(L,t)\equiv \sqrt{\left<(h-\left<h\right>)^2\right>}
\end{equation}
for the initial condition $h(x,0)=0$. In Fig. \ref{fig: FV5},
we plot $W/L^{1/2}$ against $t/L^{3/2}$
for system sizes $L=64, 256, 1024,$ and $4096$ with 
$\nu=D=1$ and $\lambda=5$ fixed. It is found that all the data
are on the same universal curve. 
The scaling function observed in the simulation is consistent with
known results. 

We note that the same procedure does not yield
a clear curve for the numerical simulations of the system with $\lambda=1$,
as shown in Fig. \ref{fig: FV1}. This is interpreted as a finite
time effect.
Concretely, in the early stage  $t \ll t_0$,
the growth of $W$ is described by the linear (equilibrium) dynamics
with $\lambda=0$, while the non-linear growth effect becomes dominant
for the late stage $t \gg t_0$. The cross-over time $t_0$ was numerically obtained \cite{Krug1997, KPZ_review_H1995, Sneppen1992} as 
\begin{equation}
    t_0 \simeq 252\nu^5 D^{-2}\lambda^{-4}.
\end{equation}
We conjecture that the cross-over time $t_0$ is related to the
cross-over length $L_0$ in our study.     

\begin{figure}[t]
      \begin{minipage}[t]{0.48\hsize}
    \centering
    \includegraphics[width=8cm]{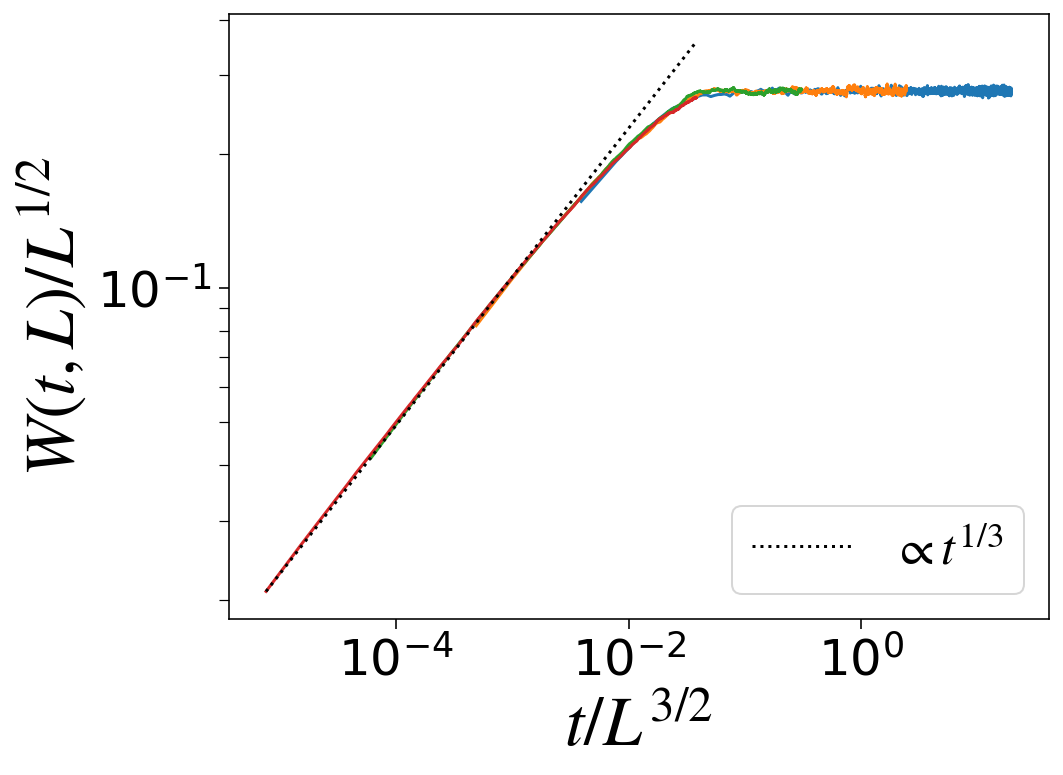}
    \caption{Dynamical scaling of the width $W$ for the KPZ interface with $\lambda=5$. The system sizes are $L=64, 256,1024$, and $4096$ from right to left. The dotted line is a guide to the eye.}
    \label{fig: FV5}
      \end{minipage}
      \hfill
      \begin{minipage}[t]{0.48\hsize}
    \centering
    \includegraphics[width=8cm]{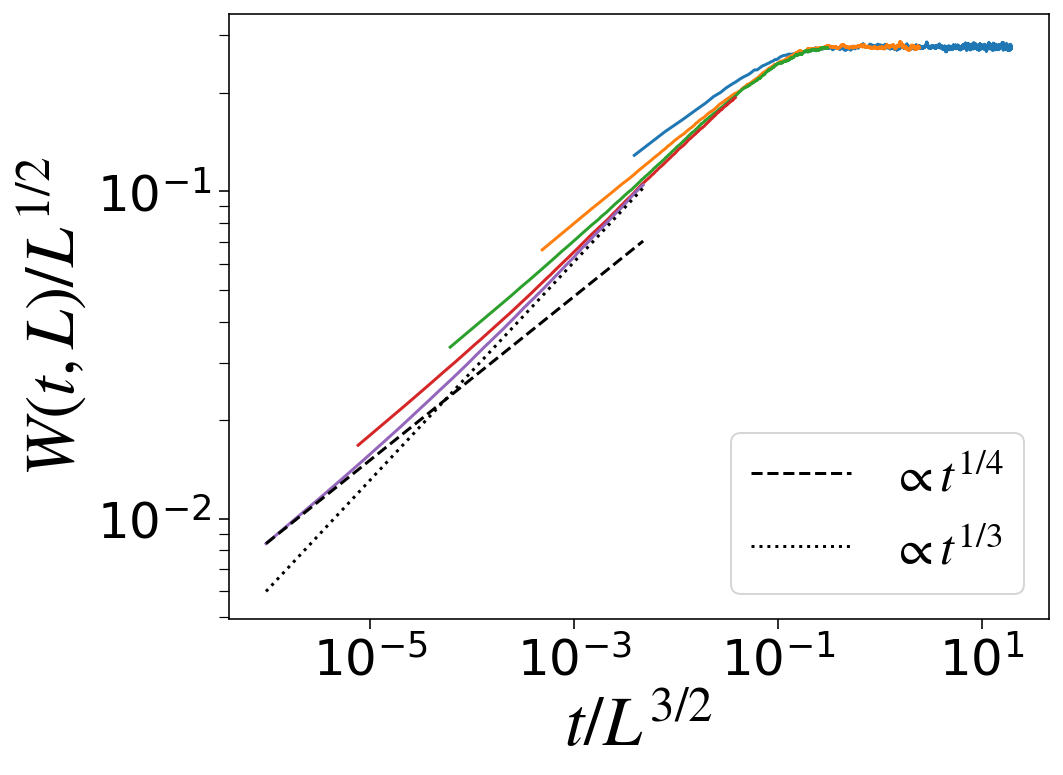}
    \caption{The same as Fig. \ref{fig: FV5}, but for $\lambda=1$. The system sizes are $L=64, 256,1024,4096$, and $16384$ from right to left. In this case, the width grows as $t^{1/4}$ (dashed line), not $t^{1/3}$ (dotted line), until non-linearity grows enough.}
        \label{fig: FV1}
      \end{minipage}
\end{figure}

\subsection{Finite mesh effect for $\kappa_{\rm eff}(L)$}
\label{ssec: Finite-size effect}

Since we fix $\Delta x=0.5$ in the discrete model, the model with small $L$ may not provide a good approximation
of the KPZ equation.  In order to study this aspect more quantitatively, in Fig. \ref{fig: L_k_lam5}, we plot $\kappa_\eff$
against $L \ge 2$ with $v_0=5$ fixed. It shows that $\kappa_\eff$ is
almost proportional to $\sqrt{L}$ for $L \ge 4$. 
However, when we plot $\kappa_\eff$ for several values of $v_0$
in Fig. \ref{fig: L_k_full}, we find that
the data for $L=2, 4,$ and $8$ are not on the one universal curve.
This means that the discrete model with $\Delta x=0.5$ is
not a good approximation of the KPZ equation with $L=2, 4,$ and $8$.
More quantitatively, we notice that the cut-off number $n_c=L/(2\Delta x)$ characterizes the cross-over length as shown in \eqref{Eq: l} and Fig. \ref{fig: l}. This means that the system with smaller $n_c$ shows a shorter cross-over length, which makes the data points shift to the right side. Therefore, the data for the systems with small $L$ are obviously contaminated by discretization effects and thus deviate from the universal curve determined in the larger systems. For this reason, we employ the data for $L \ge 16$ in the main text.

\begin{figure}[t]
      \begin{minipage}[t]{0.48\hsize}
    \centering
    \includegraphics[width=8cm]{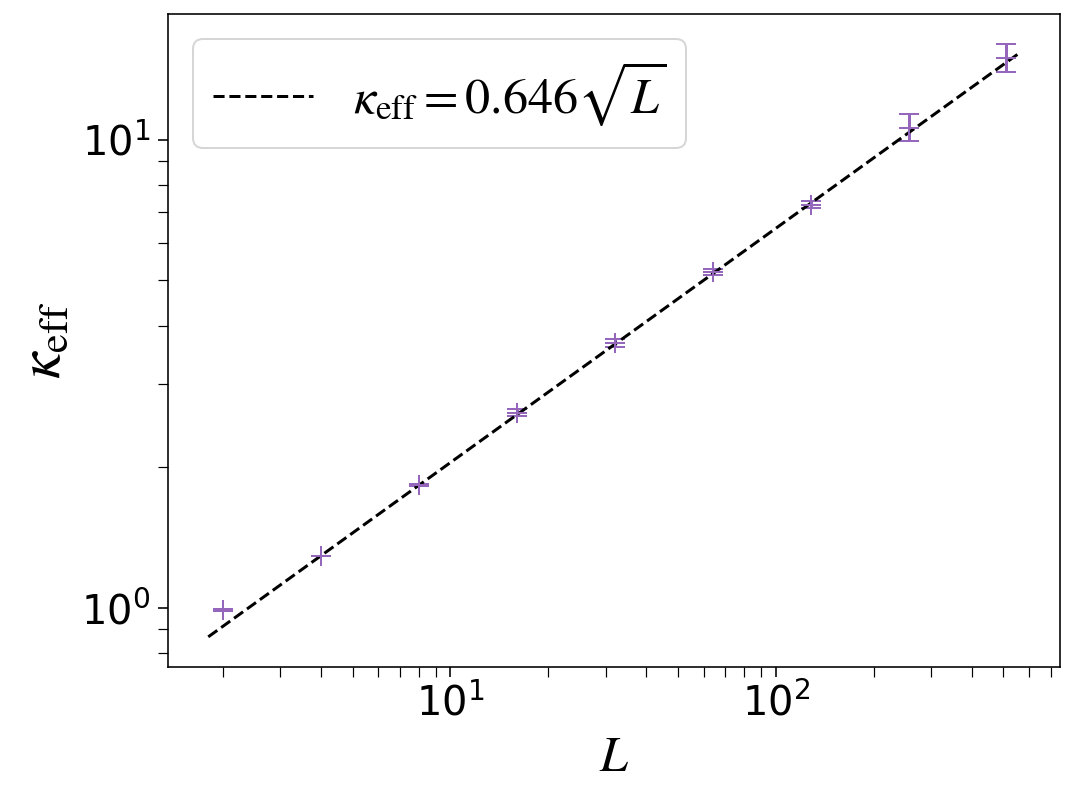}
    \caption{$\kappa_\eff$ for $L = 2, 4, 8, 16, 32, 64, 128, 256$, and $512$ from left to right.  $v_0=5$. The symbols $L \ge 4$ are on the straight line
    corresponding to $\kappa_\eff=0.646\sqrt{L}$.}
    \label{fig: L_k_lam5}
      \end{minipage}
      \hfill
      \begin{minipage}[t]{0.48\hsize}
    \centering
    \includegraphics[width=8cm]{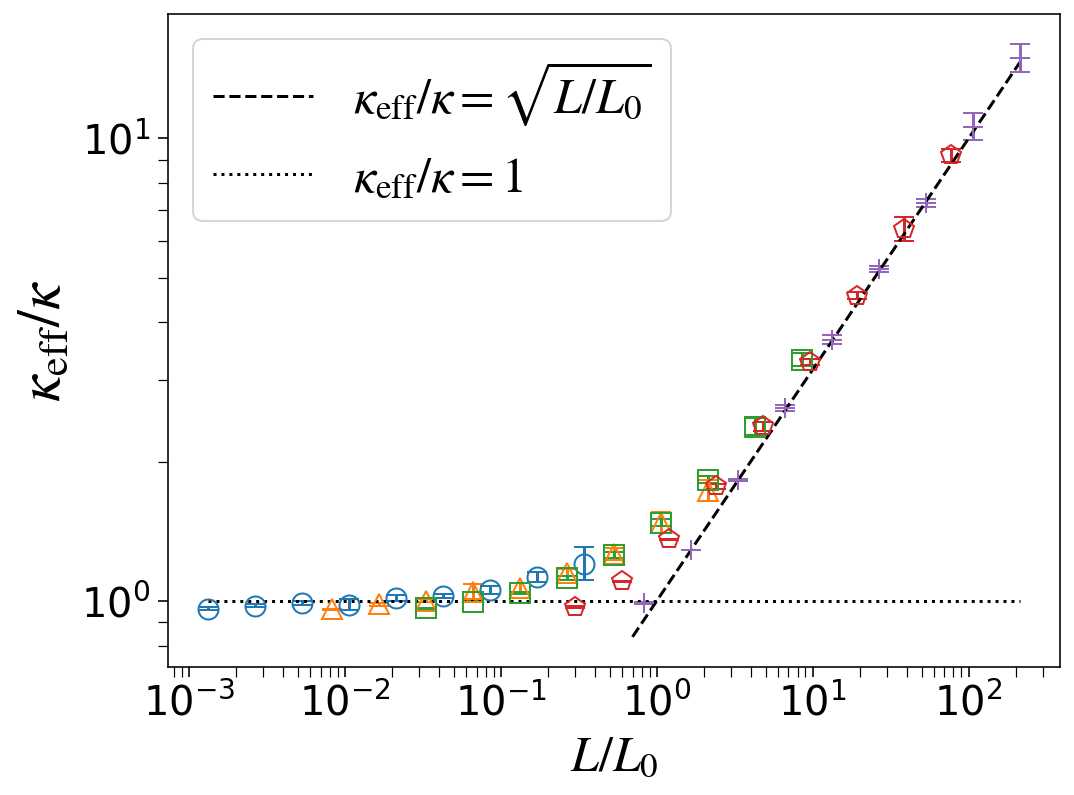}
    \caption{$\kappa_\eff$ for the same system sizes as Fig. \ref{fig: L_k_lam5} and with $v_0=0.2, 0.5, 1, 3$, and $5$ (difference in symbols represents the difference in $v_0$). $L_0$ is given by \eqref{Eq: L0} with $\ell\simeq 60$ in this figure. The symbols for the small system sizes $L=2,4$, and $8$ are not
in the universal curve.}
  \label{fig: L_k_full}
      \end{minipage}
\end{figure}

\newpage


\end{document}